\documentclass[12pt]{article}
\usepackage{amsmath,epsf,amssymb,latexsym,cite,shadow,eucal,theorem}
\usepackage[ps,dvips,matrix,arrow,frame,import,curve,color]{xy}

\newtheorem{theorem}{Theorem}
\newtheorem{prop}{Proposition}

\newtheorem{definition}{Definition}
{\theorembodyfont{\rmfamily}\newtheorem{example}{Example}}

\newif\iffigs\figstrue

\DeclareFontFamily{U}{rsf}{}
\DeclareFontShape{U}{rsf}{m}{n}{
  <5> <6> rsfs5 <7> <8> <9> rsfs7 <10-> rsfs10}{}
\DeclareMathAlphabet\Scr{U}{rsf}{m}{n}

\def\O{\Scr{O}}
\def\C{{\mathbb C}}
\def\P{{\mathbb P}}

\def\R{{\mathbb R}}
\def\Z{{\mathbb Z}}

\def\NN{{\mathbb N}}

\def\Hom{\operatorname{Hom}}

\def\Ext{\operatorname{Ext}}
\def\Tor{\operatorname{Tor}}
\def\End{\operatorname{End}}

\def\Spec{\operatorname{Spec}}

\def\Tr{\operatorname{Tr}}

\def\diag{\operatorname{diag}}

\def\lmod#1{\hbox{$#1$--{\bf mod}}}

\def\pz{\phantom{-}}

\def\CY{Calabi--Yau}

\def\cA{{\Scr A}}

\def\cP{{\Scr P}}

\def\cE{{\Scr E}}
\def\cF{{\Scr F}}
\def\cX{{\Scr X}}

\def\DC{\mathbf{D}}

\def\QED{$\quad\blacksquare$}
\def\ff#1#2{{\textstyle\frac{#1}{#2}}}

\def\mf#1{\mathfrak{#1}}

\def\poso#1{#1\save="x"!LD+<0pt,-0.5mm>;
  "x"!RD+<0pt,-0.5mm>**\dir{.}\restore}

\def\balpha{{\boldsymbol{\alpha}}}
\def\bbeta{{\boldsymbol{\beta}}}
\def\bdelta{{\boldsymbol{\delta}}}

\begin{document}

\begin{titlepage}
\begin{flushright}
DUKE-CGTP-08-01\\
June 2008
\end{flushright}
\vspace{.5cm}
\begin{center}
\baselineskip=16pt
{\fontfamily{ptm}\selectfont\bfseries\huge
D-Branes
on Toric \CY\ Varieties}\\[20mm]
{\bf\large  Paul S.~Aspinwall
 } \\[7mm]

{\small

Center for Geometry and Theoretical Physics, 
  Box 90318 \\ Duke University, 
 Durham, NC 27708-0318 \\ \vspace{6pt}

 }

\end{center}

\begin{center}
{\bf Abstract}
\end{center}
We analyze B-type D-branes on noncompact toric Calabi--Yau spaces. A
general program is presented to find a set of tilting line bundles
that yields the associated quiver and its relations. In many cases,
this set remains fixed as one moves between phases in the K\"ahler
moduli space. This gives a particularly simple picture of how the
derived category remains invariant across all phases. The
combinatorial problems involving local cohomology used to determine
the tilting set are also related to questions of $\Pi$-stability as
one moves between phases. As a result, in some cases precisely
those line bundles in the tilting set remain stable over the whole
moduli space in some sense.

\end{titlepage}

\vfil\break


\section{Introduction}    \label{s:intro}

Toric varieties form a wonderful playground in providing a large
class of algebraic varieties in which difficult questions in algebraic
geometry can be reduced to combinatorics. In the physics of string
theory, toric geometry appears in the form of gauged linear
$\sigma$-models with an abelian gauge group
\cite{W:phase,AGM:II}. While one can argue that toric geometry
certainly does not represent truly generic algebraic varieties, and thus
generic string theory vacua, one can still learn valuable lessons by
understanding this ``easy'' case first. 

A \CY\ variety cannot be both compact and toric. Here we restrict
attention to noncompact toric \CY\ varieties. B-type topological
D-branes on a \CY\ variety are represented by the bounded derived
category of coherent sheaves $\DC(X)$ (see \cite{me:TASI-D} for a
review and references). It is well-known that D-branes on noncompact
\CY\ spaces (and thus the derived category of noncompact \CY\ spaces)
are best understood in terms of quivers. This paper is yet another in
the vast literature of this topic to address the interplay of
D-branes, derived categories and quivers.

The basic tool of the analysis in this paper is to make use of
``tilting line bundles'' which are line bundles supported over the
whole noncompact space. These are the analogues of tautological line
bundles in the McKay correspondence \cite{GS-V:Mckay}. These bundles
have also played an important r\^ole in the recent work
\cite{HHP:linphase} where D-branes are analyzed directly in terms of
the gauged linear sigma model. Tilting bundles are very similar
to ``exceptional collections'' of bundles on a compact Fano variety
which have been used in this context many times (see, for example,
\cite{CFIKV:,Herz:exc,AM:delP,Herzog:2005sy}). Tilting collections
have also previously been used to analyze D-branes in some examples
\cite{BD:tilt,Brn:tilt}. The use of tilting bundles allows one to
avoid the assumption that the noncompact \CY\ is the total space of a
bundle over some compact irreducible variety. This, in turn, leads to
a picture of $\DC(X)$ that is not particularly tied to any phase.

The goal of this paper was to reduce the main questions of $\DC(X)$
and D-branes on a toric \CY\ to purely combinatorial questions and
thus solve them. We have not been completely successful in this
regard as we are unable to solve the general combinatorial
problems. However, given an analysis of many examples, a general
picture which is very pretty appears to emerge.

The basic claim is that one can have a tilting set of line bundles,
which describes a quiver and thus $\DC(X)$, which is {\em globally\/}
defined over the K\"ahler moduli space (if certain cuts are made to
avoid arbitrary monodromy). This immediately yields a very direct
picture of why $\DC(X)$ is invariant between phases. This property,
which we call ``wholesomeness'' is defined more carefully below and we
demonstrate its validity for some classes and specific examples of
toric \CY s. It is tempting to conjecture that wholesomeness is true
in all cases.

The use of tilting objects as described in this paper gives yet
another way of deriving the quiver gauge theory, complete with
superpotential, from toric data describing a singularity. Other
methods include resolving orbifolds \cite{MP:AdS} and dimers
\cite{Franco:2005sm}. The connection between dimers and the case where
a tilting object can be derived from an exceptional collection was
discussed in \cite{Hanany:2006nm}. We believe the tilting object
method described in this paper is the quickest and mathematically most
direct way of computing the quiver from the toric data but this could
be a subjective statement.

As well as describing $\DC(X)$, one would like to understand
$\Pi$-stability over the K\"ahler moduli space. This turns out to be
very closely related to the mathematics associated to finding the
tilting set. Thus, we again arrive at combinatorial problems
associated with toric geometry when addressing these questions.

In section \ref{sec:quiv-tilt-sheav} we review the basic ideas of
tilting sheaves and how the derived category of coherent sheaves is
written in terms of a quiver with relations. In section
\ref{sec:toric-cy-s} we review the algebraic geometry and commutative
algebra we require from toric geometry.

The main part of the paper is section \ref{sec:DX-generated-line}
which analyzes how one might go about finding a tilting set of line
bundles for a given \CY. The notion of wholesomeness is introduced and
various classes and examples are discussed. In section
\ref{sec:pi-stab} the relationship to $\Pi$-stability is discussed and
finally we present some concluding remarks.


\section{Quivers and Tilting Sheaves}  \label{sec:quiv-tilt-sheav}

Let $X$ be a smooth \CY\ variety over $\C$ which may be noncompact and
let $\DC(X)$ denote the bounded derived category of coherent sheaves
on $X$. As usual, if $a$ denotes a complex in $\DC(X)$ then $a[n]$
denotes the same complex shifted $n$ places to the left.  
A single coherent sheaf $\cF$ is regarded as an object in $\DC(X)$ in
terms of a complex which is zero in every position except position
zero, where it is $\cF$.
Suppose we
can find a {\em tilting sheaf\/} $M$ which is a coherent sheaf on $X$
such that\footnote{We also impose the technical requirement that $A$
  (defined below) has finite global dimension.}
\begin{enumerate}
\item  $M$ decomposes as a finite direct sum of simple sheaves
\begin{equation}
  M = P_1\oplus P_2 \oplus\ldots\oplus P_k.   \label{eq:tilt}
\end{equation}
\item Each $P_i$ satisfies
\begin{equation}
  \Ext^n(P_i,P_j)=0 \quad\hbox{for all $n>0$ and all $i,j$.}
\end{equation}
\item The collection of $P_i$'s generates the whole of $\DC(X)$. In
  other words, the smallest triangulated full subcategory of $\DC(X)$
  containing $\{P_1,\ldots,P_k\}$ is $\DC(X)$ itself.
\end{enumerate}

Let $A=\End(M)$ be the endomorphism algebra of $M$. The product rule
in $A$ is simply composition of maps of $M$ to itself. We may view
elements of $A$ as matrices whose $(i,j)$th entry is an element of
$\Hom(P_j,P_i)$. It is then clear that the
product is not, in general, commutative. $M$ has the structure of a
bimodule with a left action from $\O_X$, as it is a sheaf, and a right
action by $A$. This leads to a well-known equivalence
\cite{Baer:quiv,Rick:der,Bon:dPq} 
\begin{equation}
  \DC(X) \cong \DC(\lmod A),
\end{equation}
where $\DC(\lmod A)$ is the bounded derived category of finitely
generated left $A$-modules.\footnote{There are related statements
  such as the equivalences concerning bounded derived categories of
  sheaves with compactly supported cohomology. See, for example,
  \cite{Bridge:Z3,Berg:compact}.} This equivalence is induced by an
  adjoint pair of functors
\begin{equation}
\begin{split}
  \Hom(M,-)&:D(X) \to \DC(\lmod A)\\
  M\otimes_A -&:\DC(\lmod A) \to D(X).
\end{split}  \label{eq:FGequiv}
\end{equation}

The noncommutative algebra $A$ can be written in terms of the path
algebra of a quiver $Q$ with relations. We may associate a node to
each summand $P_i$. Then $\Hom(P_i,P_j)$ is generated, as a
vector space, by paths from node $j$ to node $i$. That is, arrows
represent {\em indecomposable\/} maps between the sheaves $P_i$.  Note
that $\End(T)$ is the path algebra of $Q^{\textrm{op}}$, the quiver
$Q$ with all arrows reversed.  An $A$-module may be identified
with a quiver representation as discussed at length in
\cite{King:th,Brn:tilt,AM:delP,Herz:ouch} for example.

Under (\ref{eq:FGequiv}) the sheaf $P_i$ in $\DC(X)$ is mapped to
$\Hom(M,P_i$). This has the interpretation of the space of all paths
starting at node $i$. With a slight abuse of notation we will
also use $P_i$ to refer to this representation of $Q$. Let
$e_i$ denote the trivial path of length zero beginning and ending at
node $i$. The representation of $Q$ given by $P_i$ may then be written
$Ae_i$.

\begin{example}
  The classic example is $X=\P^n$ due to Beilinson \cite{Bei:res}. For
  example, if $X$ is $\P^2$, with homogeneous coordinates
  $[x_0,x_1,x_2]$, we may put $P_i=\O(i)$ for $i=0,1,2$. This yields a
  quiver
\begin{equation}
\begin{xy} <1.0mm,0mm>:
  (0,0)*{\circ}="a",(20,0)*{\circ}="b",(40,0)*{\circ}="c",
  (0,-4)*{v_0},(20,-4)*{v_1},(40,-4)*{v_2}
  \ar@{<-}@/^3mm/|{a_0} "a";"b"
  \ar@{<-}|{a_1} "a";"b"
  \ar@{<-}@/_3mm/|{a_2} "a";"b"
  \ar@{<-}@/^3mm/|{b_0} "b";"c"
  \ar@{<-}|{b_1} "b";"c"
  \ar@{<-}@/_3mm/|{b_2} "b";"c"
\end{xy}  \label{eq:egP2}
\end{equation}
Both $a_i$ and $b_i$ correspond to multiplication by $x_i$.
This yields relations $a_ib_j=a_jb_i$ for all $i,j$. 
\end{example}

In the above example we have a {\em directed\/} quiver without loops
and $A$ is finite-dimensional. In this case, the tilting set
$\{P_0,P_1,\ldots\}$ form an {\em exceptional collection}. In this
paper we will be more concerned with cases where the quiver has loops
and thus $A$ is infinite-dimensional.

\begin{example}
Now consider the total space of the line bundle with $c_1=-3$ over
$\P^2$. This is a noncompact \CY\ threefold. Again we put
$P_i=\O(i)$ for $i=0,1,2$, but now these line bundles have noncompact
support. The quiver looks like
\begin{equation}
\begin{xy} <1.0mm,0mm>:
  (0,0)*{\circ}="a",(40,0)*{\circ}="b",(20,28)*{\circ}="c",
  (0,-4)*{v_0},(40,-4)*{v_1},(20,32)*{v_2}
  \ar@{<-}@/^3mm/|{a_0} "a";"b"
  \ar@{<-}|{a_1} "a";"b"
  \ar@{<-}@/_3mm/|{a_2} "a";"b"
  \ar@{<-}@/^3mm/|{b_0} "b";"c"
  \ar@{<-}|{b_1} "b";"c"
  \ar@{<-}@/_3mm/|{b_2} "b";"c"
  \ar@{<-}@/^3mm/|{c_0} "c";"a"
  \ar@{<-}|{c_1} "c";"a"
  \ar@{<-}@/_3mm/|{c_2} "c";"a"
\end{xy}  \label{eq:egO(-3)}
\end{equation}
The extra maps $c_i$ are given by multiplication by $px_i$ where $p$
is the coordinate in the fibre direction.  This gives relations
$a_ib_j=a_jb_i, b_ic_j=b_jc_i, c_ia_j=c_ja_i$ for all $i,j$.
\end{example}

Consider the {\em center\/} $Z(A)$ of the path algebra $A$ of
(\ref{eq:egO(-3)}). The rings $\Hom(P_i,P_i)$ are isomorphic for all
$i$ to some ring which we denote $R$. Elements of $Z(A)$ are then
given by matrices of the form $\diag(r,r,r)$ for $r\in R$ and thus
$Z(A)$ is isomorphic to $R$. $R\cong\Hom(P_0,P_0)$ is then generated
as a ring by
\begin{equation}
  x_{ijk} = a_i b_j c_k, \quad\hbox{where $i\leq j\leq k$.}
\end{equation}
Now let $G=\Z_3$, generated by $g$, act on $(u_0,u_1,u_2)$ by
\begin{equation}
  g:(u_0,u_1,u_2) \mapsto (\omega u_0,\omega u_1,\omega u_2),
\end{equation}
where $\omega$ is a nontrivial cube root of unity. It is easy to
see that $R$ is isomorphic to the $G$-invariant part of the polynomial
ring $\C[u_0,u_1,u_2]$ by putting $x_{ijk} = u_iu_ju_k$. That is,
\begin{equation}
  \Spec Z(A) = \C^3/\Z_3.
\end{equation}

This is a typical example of a {\em noncommutative resolution\/} in
the sense of \cite{Berenstein:2001jr,bergh04:nc,ginzburg06:CYa}. The
singular variety $\C^3/\Z_3$ has a crepant ``resolution'' by the
noncommutative algebra $A$.


\section{Toric \CY s}   \label{sec:toric-cy-s}

First we review a standard construction in toric geometry.  Let $N$ be
a lattice of rank $d$. Let $\cP$ be a convex polytope in $N\otimes\R$
such that the vertices of the convex hull lie in $N$. Furthermore, we
demand that $\cP$ lies in a hyperplane of $N\otimes\R$ such that the
coordinates of any point in $\cP$ may be written $(1,\ldots)$. Let
$\cA$ denote the set of points $\cP\cap N$ and let $n$ denote the
number of elements of $\cA$.

The coordinates of the points of $\cA$ form a $d\times n$ matrix
defining a map $A:\Z^{\oplus n}\to N$ which we assume is surjective. We form
an exact sequence
\begin{equation}
\xymatrix@1@M=2mm{
  0\ar[r]&L\ar[r]&\Z^{\oplus n} \ar[r]^A&N \ar[r]& 0,
} \label{eq:LZN}
\end{equation}
where $L$ is the ``lattice of relations'' of rank $r=n-d$. Dual to
this we write
\begin{equation}
\xymatrix@1@M=2mm{
  0\ar[r]&M\ar[r]&\Z^{\oplus n}\ar[r]^\Phi&D\ar[r]& 0,
} \label{eq:M-D}
\end{equation}
where $\Phi$ is the $r\times n$ matrix of ``charges'' of the points in
$\cA$. By our hyperplane condition, each row of $\Phi$ sum to zero.

Let
\begin{equation}
  S = \C[x_1,\ldots,x_n].
\end{equation}
The matrix $\Phi$ gives an $r$-fold multi-grading to this ring. In
other words, we have a $(\C^*)^r$ torus action:
\begin{equation}
  x_i \mapsto \lambda_1^{\Phi_{1i}}\lambda_2^{\Phi_{2i}}\ldots
              \lambda_r^{\Phi_{ri}} x_i,
\end{equation}
where $\lambda_j\in\C^*$. Let $R$ be the $(\C^*)^r$-invariant
subalgebra of $S$. The algebra $S$ then decomposes into a sum of
$R$-modules labeled by their $r$-fold grading:
\begin{equation}
  S = \bigoplus_{\balpha\in D} S_{\balpha},
\end{equation}
where $D\cong\Z^{\oplus r}$ from (\ref{eq:M-D}) and $R=S_0$.
As usual we denote a shift in grading by parentheses, i.e.,
$S(\balpha)_\bbeta=S_{\balpha+\bbeta}$.

Let $X_0=\Spec R$. That is, $X_0$ is the toric variety associated to
the fan consisting of the single cone over $\cP$. $X_0$ is then a
noncompact (typically) singular \CY\ variety. We would like to find a
non-commutative crepant resolution of $X_0$. This problem was solved
completely in the last section of \cite{bergh04:nc} for the case
$r=1$. We would like to examine the general case.

It is well-known in toric geometry that a (partial) crepant
desingularization of $X_0$ is given by a simplicial decomposition of
the point set $\cA$. In order that this desingularization be K\"ahler
we also impose that the simplicial decomposition be ``regular''
\cite{OP:convex}. 
This simplicial decomposition may, or may not,
include points in the interior of the convex hull of $\cA$. We refer
to a choice of simplicial decomposition as a ``phase''. $X_0$
corresponds to a phase itself if and only if the convex hull of $\cA$
is a simplex. A phase corresponds to a complete resolution if each
simplex has volume one (in the natural normalization by
$(d-1)!$). Otherwise a phase has orbifold singularities.

To each phase we associate the ``Cox ideal'' defined in \cite{Cox:} as
follows. 
\begin{definition}
Let $\Sigma=\{\sigma_1,\sigma_2,\ldots\}$ denote the set of
simplices. If $\sigma$ is a simplex, we say $i\in\sigma$ if the $i$th
element of $\cA$ is a vertex of $\sigma$. Then
\begin{equation}
  B_\Sigma =
  \left(\prod_{i\not\in\sigma_1}x_i,\prod_{i\not\in\sigma_2}x_i, 
  \ldots\right).
\end{equation}
\end{definition}
Clearly $B_\Sigma$ is a square-free monomial ideal in $S$.

\begin{definition}
Let $V(B_\Sigma)$ denote the subvariety of $\C^n$ given by
$B_\Sigma$. Then
\begin{equation}
  X_\Sigma = \frac{\C^n - V(B_\Sigma)}{(\C^*)^r}.
\end{equation}
\end{definition}

Cox \cite{Cox:} shows that there is a correspondence between
finitely-generated graded $S$-modules and coherent sheaves on a smooth
$X_\Sigma$ which follows the usual correspondence between sheaves and
projective varieties as in chapter II.5 of \cite{Hartshorne:}. If $U$
is an $S$-module, we denote $\widetilde U$ as the corresponding
sheaf. $\widetilde U$ is zero as a sheaf if and only if $U$ is killed
by some power of $B_\Sigma$. This yields
\begin{prop}  \label{prop:Cox}
Assume $X_\Sigma$ is a smooth toric variety. Then
\begin{equation}
  \DC(X_\Sigma) = \frac{\DC(\mathrm{gr-}S)}{T_\Sigma},  \label{eq:stack}
\end{equation}
where $\DC(\mathrm{gr-}S)$ is the bounded derived category of
finitely-generated multigraded $S$-modules and $T_\Sigma$ is the full
subcategory generated by modules killed by a power of $B_\Sigma$. This
quotient of triangulated categories is as in \cite{BO:flop}.
\end{prop}

If $\Sigma$ does not consist of simplices of volume one, $X_\Sigma$
will have orbifold singularities and proposition \ref{prop:Cox} will
not hold. However, if we view the resulting $X_\Sigma$ as a smooth
{\em stack\/}\footnote{There is also a notion of a ``toric stack''
  where extra data is added to denote a lattice point in $N$ lying on
  each one-dimensional ray in the fan (see, for example,
  \cite{BCS:toric}). We are not using this technology here.
  There is also a related notion of boundary divisor that has been
  analyzed in the context of the derived category and toric geometry
  in \cite{Kaw:toric}.}  then the proposition is valid (by an argument
essentially given in \cite{Orlov:mfc}). So, the question we need to
address is whether D-branes on an orbifold are described by the
derived category of a variety or a stack. It has been argued in
\cite{Pantev:2005wj} that stacks are the correct language for
D-branes. Indeed, one may view \cite{HHP:linphase} as a linear
$\sigma$-model demonstration of this idea as that paper shows that the
D-branes are given by the quotient (\ref{eq:stack}). So, from now on
we will assume that the above proposition holds in any phase and we no
longer need assume that $X_\Sigma$ is smooth.


\section{$\DC(X)$ Generated by Line Bundles}  \label{sec:DX-generated-line}

\subsection{Tilting Line Bundles}

To a bounded derived category $\DC(X_\Sigma)$ we may associate the
more crude ``Grothendieck group''. This is simply the abelian group
generated by all objects in $\DC(X)$ modulo relations generated by
distinguished triangles. This is also the K-theory group of $X_\Sigma$
if $X_\Sigma$ is a smooth manifold and so measures ``D-brane
charge''. We denote the rank of the Grothendieck group by $T$.
Clearly at least $T$ objects are needed to generate the derived category.

If $X_\Sigma$ is a crepant smooth resolution, can we find a module of
the form
\begin{equation}
  M = S(\balpha_1)\oplus S(\balpha_2)\oplus \ldots\oplus S(\balpha_T),
\end{equation}
playing the r\^ole of (\ref{eq:tilt}) to provide a tilting sheaf?
That is, can we find a tilting sheaf that is a sum of line bundles?
This is closely related to the question of whether we can always find
strong exceptional collections of line bundles for toric varieties as
proposed by King \cite{King:conj}. Even though this is known not to be
the case in general \cite{HillePerling:notKing}, it may well still be
true in the ``nef-Fano'' case which corresponds to a fan over a convex
set \cite{BorisHua:King} which is the case at hand in this paper.

The first condition that we require is $\Ext^k_{X_\Sigma}(\widetilde
S(\balpha_i),\widetilde S(\balpha_j))=0$ for all $k>0$ and all
$i,j=1,\ldots, T$. That is,
\begin{equation}
  H^k(X_\Sigma,\widetilde S(\balpha_j-\balpha_i))=0. \label{eq:H0}
\end{equation}

The cohomology of line bundles on a toric variety is easily computed
via local cohomology \cite{Grot:localcoh,EMS:ToricCoh}. See
also \cite{Herzog:2005sy} for an account in the physics
literature. First define 
\begin{equation}
  H^i_*(\widetilde S) = \bigoplus_{\bdelta\in D}H^i(\widetilde
  S(\bdelta)).  \label{eq:H*}
\end{equation}
Now $H^i_*(\widetilde S)$ has the structure of a graded $S$-module as
can be seen as follows. The direct sum in (\ref{eq:H*}) decomposes
$H^i_*(\widetilde S)$ into its graded parts. Suppose $s\in
S_\bbeta$. Then we have a degree zero map $S(\bdelta)\to
S(\bdelta+\bbeta)$ given by multiplication by $s$. Then, applying the
corresponding functors, this extends to a map $H^i(\widetilde S(\bdelta))\to
H^i(\widetilde S(\bdelta+\bbeta))$. 

If $I$ is an ideal in $S$ then we denote {\em local cohomology\/} by
$H^i_I$. For more information on local cohomology we refer to
\cite{BS:LocalCoh}.

Then we have
\begin{prop}
\begin{equation}
\xymatrix@1@M=2mm{
  0\ar[r]& H^0_{B_\Sigma}(S)\ar[r]&S\ar[r]&H^0_*(\widetilde S)\ar[r]&
     H^1_{B_\Sigma}(S)\ar[r]&0,}
\end{equation}
and
\begin{equation}
  H^k_*(\widetilde S)\cong H^{k+1}_{B_\Sigma}(S)\quad\textrm{for~} k>0.
\end{equation}
\end{prop}

So, following (\ref{eq:H0}) we want to find elements $\bdelta\in D$
such that $H^k_{B_\Sigma}(S)_\bdelta=0$ for all $k\geq2$. Actually we
will impose a slightly stronger condition to include $k=0$ and 
1:\footnote{Actually $H^0_{B_\Sigma}(S)$ is always zero.}
\begin{definition}
  A vector $\bdelta\in D$ is called ``$B_\Sigma$-acyclic'' if the local
  cohomology groups $H^k_{B_\Sigma}(S)_\bdelta$ vanish for all $k$.
\end{definition}
Therefore a $B_\Sigma$-acyclic vector $\bdelta$ yields
$H^0(\widetilde S(\bdelta))=S_\bdelta$.

\subsection{Computing Local Cohomology}

Computing local cohomology is particularly easy when the ideal is a
monomial ideal. We review the details of the construction of Musta{\c
  t}{\v a} \cite{Must:LocalCoh} as we will need them later in this
paper.

$S$ has an $r$-fold grading given by the matrix of charges $\Phi$. It
also has an $n$-fold ``fine'' grading where we simply assign $x_i$ a
grading of $(0,0,\ldots,0,1,0,\ldots,0)$ where the 1 appears in the
$i$th position. The matrix $\Phi$ can then be viewed as a map from the
lattice of fine grading to the $D$-lattice. We will use non-bold
letters $\alpha,\ldots$ for fine grading vectors.

Given a square-free monomial ideal $B$, we denote the Alexander dual
of $B$ by $B^\vee$. We refer to chapter one of \cite{MS:combcomm} for
a nice account of Alexander duality. Note that the Alexander dual of
the Cox ideal of a toric variety is the {\em Stanley--Reisner ideal}
$I_\Sigma$. That is, $I_\Sigma=B_\Sigma^\vee$ is generated by monomials of the
form $x_ix_jx_k\ldots$ where $i,j,k\ldots$ are {\em not\/} the
vertices of any simplex in the triangulation specified by $\Sigma$.

Consider a minimal finely-graded free resolution of $B^\vee$:
\begin{equation}
\xymatrix@1@M=2mm{
  \cdots\ar[r]&F_2\ar[r]&F_1\ar[r]&F_0\ar[r]&B^\vee\ar[r]&0,
} \label{eq:SRres}
\end{equation}
where
\begin{equation}
  F_i = \bigoplus_{\alpha\in\NN^{\oplus n}} S(-\alpha)^{\oplus
    b_{i,\alpha}}.
\end{equation}
The numbers $b_{i,\alpha}$ are known as {\em graded Betti numbers} and
can also be written
\begin{equation}
  b_{i,\alpha} = \dim\Tor_i^S(B^\vee,\C)_\alpha,  \label{eq:betti}
\end{equation}
where $\C$ is an $S$-module annihilated by any $x_i$. One may
similarly define $D$-graded Betti numbers $b_{i,\balpha}$ for
$\balpha\in D$.

Note that the $\alpha$'s giving rise to nonzero finely-graded Betti
numbers are ``binary'' vectors in the sense that they a lists of 0's
and 1's. Let $\Xi$ be a map from $\Z^{\oplus n}$ to $\{0,1\}^n$ which
replaces non-negative numbers by 0 and negative numbers by 1.  One then
has \cite{Must:LocalCoh}
\begin{theorem}
  $H^i_B(S)_\delta$ is nonzero for some $i$ if and only if there is a
  nonzero Betti number $b_{k,\alpha}$ for some $k$ such that
  $\Xi(\delta)=\alpha$.
\end{theorem}

We therefore have an algorithm for finding valid vectors $\bdelta\in
D$ such that the local cohomology groups $H^i_B(S)_\bdelta$ vanish for
all $i$:
\begin{itemize}
\item For each nonzero Betti number $b_{k,\alpha}$, take the
corresponding orthant of $\Z^{\oplus n}$ that maps via $\Xi$ to
$\alpha$.
Project this orthant to $D$ via $\Phi$ and remove the resulting
  vectors from consideration.
\item The remaining vectors in $D$ satisfy the desired acyclic condition.
\end{itemize}
 
Once we have found the set of acyclic $\bdelta$ vectors we may then
try to find a choice of $T$ vectors $\{\balpha_i\}$ such that
$\balpha_i-\balpha_j$ is an acyclic vector for all $i,j$. Clearly given
such of a choice of $\{\balpha_i\}$ one may find another valid set by
shifting all the gradings by some fixed vector or by permuting the
$\balpha$'s. We will refer to such a change in $\{\balpha_i\}$ as {\em
  trivial\/}. 

\subsection{Wholesomeness}

Let $\{\balpha_i\}$ denote a set of $T$ vectors in $D$ such that all pairwise
differences are acyclic.

\begin{definition}
A given $X_\Sigma$ (or the associated point set $\cA$) will be said to
be ``wholesome'' if all the following conditions are met:
\begin{enumerate}
  \item The number of acyclic $\bdelta$'s need not be finite but the
    number of acyclic $\bdelta$'s such that $-\bdelta$ is also
    acyclic is finite.
  It follows that the number of choices (up to trivial
    transformations) of $\{\balpha_i\}$ is finite.
  \item $\{\balpha_i\}$ is maximal in the sense that no further
    vectors may be added such that all pairwise differences are
    acyclic.
  \item There are no nontrivial relations (in the form of a
    distinguished triangle of complexes) between the
    $\widetilde S(\balpha_i)$'s in $\DC(X_\Sigma)$.
  \item The $\widetilde S(\balpha_i)$'s generate $\DC(X_\Sigma)$ and so
    the sum of the $\widetilde S(\balpha_i)$'s is a tilting sheaf.
  \item $\{\balpha_i\}$ can be chosen to be identical
    in all phases. That is, it depends only on the choice of $\cA$ and
    not the triangulation $\Sigma$.
\end{enumerate}
\end{definition}

At first sight one might consider these conditions to be rather
stringent, especially the last one. Surprisingly, however, all the
examples we have considered in dimension three appear to be wholesome
and it is fairly tempting to speculate that wholesomeness is guaranteed for
any point set $\cA$ in this case. We will give a counterexample in
dimension 5 later.

Note that our stronger condition that $H^1_{B_\Sigma}(S)_\bdelta$
vanish, in addition to the higher cohomologies, is necessary in many
examples for wholesomeness to be true.

\begin{theorem}
Wholesomeness condition 1 is always true.
\end{theorem}

To prove this it is useful to describe the choice of triangulations
$\Sigma$ in terms of the {\em toric ideal\/} $I_\cA$ introduced by
Sturmfels \cite{Sturm:Grob}. Let $v=(v_1,\ldots,v_n)$ be a vector in
the kernel of $A$ in (\ref{eq:LZN}). Let $v=v_+-v_-$ where $v_\pm$ has
only non-negative coordinates and let $p_+$ be the subset of
$\{1,\ldots,n\}$ such that $i\in p_+$ when $v_i>0$. Similarly let
$p_-$ be the subset for which $v_i$ is negative. We then associate to
$v$ the binomial
\begin{equation}
  x^{v_+} - x^{v_-} \in S. \label{eq:tI}
\end{equation}
Here we have used the standard notation $x^v=\prod_i x_i^{v_i}$.
The ideal $I_\cA$ is then defined as the ideal in $S$ generated by
such binomials for all choices of vectors in the kernel of $A$.

Now, given any term ordering $\prec$ (see, for example,
\cite{CoxLittleO}) we may compute the initial ideal
$\textrm{in}_\prec(I_\cA)$. Sturmfels \cite{Sturm:Grob} then argued
that the set of possible initial ideals obtained by varying $\prec$
maps surjectively to regular triangulations $\Sigma$ of
$\cA$. This map is given simply by
\begin{equation}
  \sqrt{\textrm{in}_\prec(I_\cA)} = I_\Sigma,
\end{equation}
where $I_\Sigma$ is the Stanley--Reisner ideal of the triangulation.

Fix a term-ordering $\prec$ and thus a triangulation $\Sigma$. Let $m$
be one of the monomial generators of $I_\Sigma$. There is then a
primitive binomial of the form (\ref{eq:tI}) where the
support\footnote{That is, the set of elements $i\in\{1,\ldots,n\}$
  such that $x_i$ divides $m$.} of $m$
equals $p_+$. Let $v=(v_1,\ldots,v_n)$ denote the associated vector in
the kernel of $A$.  We know from the resolution (\ref{eq:SRres}) that
we have a corresponding nonzero Betti number $b_{0,\alpha}$. The
location of the 1's in $\alpha$ is precisely $p_+$ which, in turn, is
precisely the location of the positive numbers in $v$.

for $v=v_+-v_-$, we define $N_v$ as the sum of the coordinates of
$v_+$, which is equal to negative the sum of the coordinates of $v_-$
by our assumption that the rows of $\Phi$ sum to zero.

Since $v$ corresponds to a vector in the kernel of $A$, it is the
image of a vector $\mathbf{v}_{m,\Sigma}$ in $L$ from (\ref{eq:LZN}).
It follows that the set to be excluded from consideration
\begin{equation}
H_{m,\Sigma}=\Phi(\Xi^{-1}(\alpha))\subset D,
\end{equation}
will satisfy
$(\bdelta,\mathbf{v}_{m,\Sigma})\leq -N_v$ for all  $\bdelta\in
H_{m,\Sigma}$, where $(,)$ is the natural pairing between $L$ and $D$.

We may now choose other generators, $m$, of $I_\Sigma$ to remove
further regions from consideration for acyclicity. These $m$'s produce
vectors $\mathbf{v}_{m,\Sigma}$ that span all of $D\otimes \R$. This
latter statement follows from the fact that $I_\cA$ defines a variety
of dimension $d$ \cite{Sturm:Grob} and that the deformation of $I_\cA$
to $I_\Sigma$ is flat \cite{Eis:CA} and thus not
dimension-changing. Therefore, the space of allowed acyclic vectors in
$D$ does not contain a complete line passing through the origin.  \QED

\subsection{$r$=1}    \label{ss:r=1}

A particularly easy case is when $r=1$ which was analyzed in
\cite{Kawa:flip,bergh04:nc} which we essentially follow. It was also
studied in terms of the gauged linear sigma model in \cite{HHP:linphase}.  
\begin{theorem}   \label{th:r=1}
Wholesomeness is always true for $r=1$.
\end{theorem}

In this case the toric ideal $I_\cA$ has a single generator
$m_+-m_-$. We therefore have two phases $\Sigma_{\pm}$ given by an
initial ideal $(m_+)$ or $(m_-)$. Suppose $\Sigma_+$ corresponds to a
vector $v=(v_1,\ldots,v_n)$ which generates the one-dimensional kernel
of $A$. We know the $v_i$'s sum to zero and so
\begin{equation}
  \sum_{i\in p_+} v_i = -\sum_{i\in p_-}v_i = N,
\end{equation}
for some positive integer $N$.

It follows that the range of allowed elements of $D$ for which we have
nontrivial local cohomology is given by $\bdelta\in D$ for
which
\begin{equation}
  (\bdelta,\mathbf{v})<-N,
\end{equation}
where $\mathbf{v}$ generates $L$. Since $r=1$, the vector $\bdelta\in
D$ is specified by a single integer. 

Obviously, therefore, the set of tilting objects can be chosen to be
\begin{equation}
  S,S(1),S(2),\ldots,S(N-1).
\end{equation}
The same result is true for $\Sigma_-$. Thus property 5 is
satisfied.

Now consider the Koszul resolution of $S/B_{\Sigma_+}$, where
$B_{\Sigma_+}$ is the ideal $(x_{i_1},x_{i_2},\ldots)$ with
$p_+=\{{i_1},{i_2},\ldots\}$.
\begin{equation}
\xymatrix@1@M=2mm{
S(-N)\ar[r]&\oplus_{i\in p_+}S(-N+\Phi_{1i})\ar[r]&\ldots
\ar[r]&\oplus_{i\in p_+}S(-\Phi_{1i})\ar[r]&
  S\ar[r]&{\displaystyle \frac{S}{B_{\Sigma_+}}}.
}  \label{eq:Koz1}
\end{equation}
In the quotient triangulated category $\DC(X_{\Sigma_+})$
in (\ref{eq:stack}), the object $S/B_{\Sigma_+}$ is obviously in
$T_{\Sigma_+}$. It follows that we have two isomorphisms in
$\DC(X_{\Sigma_+})$:
\def\wS{\widetilde S}
\begin{equation}
\begin{split}
\wS(-N) &\cong 
\left(\xymatrix@1{\poso{\oplus_{i\in p_+}\wS(-N+\Phi_{1i})}\ar[r]&\ldots
\ar[r]&\oplus_{i\in p_+}\wS(-\Phi_{1i})\ar[r]&\wS}\right)\\
\wS &\cong 
\left(\xymatrix@1{\wS(-N)\ar[r]&\oplus_{i\in p_+}\wS(-N+\Phi_{1i})\ar[r]&\ldots
\ar[r]&\poso{\oplus_{i\in p_+}\wS(-\Phi_{1i})}}\right)
\end{split}
\end{equation}
In the above, the dotted line represents position 0 in the complex.
By using these isomorphisms and their grade-shifted counterparts, any
sheaf of the form $\wS(k)$ where $k<0$ or $k\geq N$ can be rewritten
in terms of bounded complexes using the basic set
$\wS,\wS(1),\wS(2),\ldots,\allowbreak\wS(N-1)$. Since any
finitely-generated $S$-module has a finite free resolution, it follows
that {\em any\/} such module can be written in terms of this basic
set. That is, $\DC(X_{\Sigma_+})$ is generated by these $N$
sheaves. So
\begin{equation}
  \wS\oplus \wS(1)\oplus\wS(2)\oplus\ldots\oplus\wS(N-1),
\end{equation}
is a tilting sheaf and we have proven property 4.

Now we shall prove property 3. First we need
\begin{prop}   \label{prop:gen1}
  Let $B$ be an ideal of $S$ such that $S/B$ is a regular ring. Let
  $M$ be a finitely-generated graded $S$-module that is annihilated by some
  power of $B$. Then $M$ is in the full triangulated subcategory of
  $\DC(\mathrm{gr-}S)$ generated by $S/B$ (and its grade shifts).
\end{prop}
Suppose $M$ is annihilated by $B^N$. Consider the following short
exact sequence:
\begin{equation}
\xymatrix@1@M=2mm{
0\ar[r]&BM\ar[r]&M\ar[r]&M'\ar[r]&0.
}
\end{equation}
Now $M'$ is annihilated by $B$ and is therefore an $(S/B)$-module. The
regularity condition then guarantees that $M'$ has a finite free
resolution in terms of sums of $S/B(r)$ for any grade shift $r$. The
module $BM$ is annihilated by $B^{N-1}$. Thus we prove the proposition
by induction. \QED

\medskip

So we arrive at the conclusion that $T_{\Sigma_+}$ is generated by
$S/B_{\Sigma_+}$. It follows that when performing the quotient
(\ref{eq:stack}), we need only consider triangles involving
$S/B_{\Sigma_+}$ (and its translations and shifts in grading). The
{\em only\/} relations between the $S(n)$'s are then given by the
triangles coming from the Koszul resolution (\ref{eq:Koz1}). The
Grothendieck group of $\DC(X_{\Sigma_+})$ is therefore $\Z^{\oplus
  N}$. That is, $T=N$, proving property 2, and, therefore, property 3.

Obviously the analysis for $X_{\Sigma_-}$ is identical to
$X_{\Sigma_+}$. The concludes the proof of theorem~\ref{th:r=1}. \QED

\subsection{The conifold and suspended pinch point}  \label{ssec:conifold}

One of the simplest examples is the conifold which was the principal
example of a noncommutative resolution studied by Van der Bergh
\cite{bergh:flop,bergh04:nc}. Here $n=4$ and $d=3$ (putting us in the
$r=1$ case) and the 4 points in $\cA$ form a square. Put
$S=\C[x,y,z,w]$ where the respective charges of these 4 variables are
given by
\begin{equation}
  \Phi = \begin{pmatrix}1&1&-1&-1\end{pmatrix}.
\end{equation}
The resulting toric variety is the conifold. The two resolutions,
related by a flop, are given by dividing the square $\cA$ into 2
triangles in two different ways.

Applying the results of the previous section, we have $N=2$ and a
tilting collection $\{S,S(1)\}$. The quiver is given by
\begin{equation}
\begin{xy} <1.0mm,0mm>:
  (0,0)*{\circ}="a",(30,0)*{\circ}="b",
  (0,-4)*{v_0},(30,-4)*{v_1}
  \ar@{<-}@/^4mm/|{x} "a";"b"
  \ar@{<-}@/^2mm/|{y} "a";"b"
  \ar@{<-}@/^2mm/|{z} "b";"a"
  \ar@{<-}@/^4mm/|{w} "b";"a"
\end{xy}  \label{eq:conquiv}
\end{equation}
The relations can be immediately read from this diagram given that $S$
is a {\em commutative\/} algebra, even if the path algebra of the
quiver isn't. In this case the relations are given by $xzy=yzx$,
$xwy=ywx$, etc.  It follows that the superpotential for this theory is
given by $\Tr(xzyw - yzxw)$ \cite{AF:superq}.

As another application we give the suspended pinch point of
\cite{MP:AdS} which has $r=2$. This has $S=\C[x,y,z,u,v]$ with charge
matrix given by
\begin{equation}
  \Phi = \begin{pmatrix}1&-2&1&0&\pz0\\0&-1&1&1&-1\end{pmatrix}.
\end{equation}
A tilting set is given by $\{S,S(0,1),S(1,1)\}$. The resulting quiver
is
\begin{equation}
\begin{xy} <1.0mm,0mm>:
  (20,0)*{\circ}="a",(0,28)*{\circ}="b",(40,28)*{\circ}="c",
  (20,-4)*{S(0,0)},(-8,28)*{S(0,1)},(48,22)*{S(1,1)}
  \ar@{<-}@/^2mm/|{u} "a";"b"
  \ar@{<-}@/^2mm/|{v} "b";"a"
  \ar@{<-}@/^2mm/|{x} "b";"c"
  \ar@{<-}@/^2mm/|{yz} "c";"b"
  \ar@{<-}@/^2mm/|{xy} "c";"a"
  \ar@{<-}@/^2mm/|{z} "a";"c"
  \ar@{<-}@(ur,dr)|{uv} "c";"c"
\end{xy}  \label{eq:sppquiv}
\end{equation}
and the relations (and thus superpotential) can be easily deduced from
the expressions on the arrows as in the conifold above. This example
has 5 phases and is wholesome.

Amongst the numerous ways of computing quivers and superpotentials
(see, for example
\cite{KW:coni,MP:AdS,AK:ainf,FHH:toric,Franco:2005sm}, and
\cite{Herzog:2005sy} to which it is closest) this method
seems to be the mathematically most direct.

\subsection{An orbifold example with $r>1$}

To try and go systematically beyond the case $r=1$ we consider the relatively
simple situation of an orbifold.  Suppose the convex hull of the point
set $\cA$ is a simplex.  This simplest phase to address is the
``unresolved phase'' which refers to the triangulation of $\cA$ that
has just one simplex and all points other than the vertices of the
convex hull are ignored. It is interesting to ask if such a phase is
wholesome (omitting, of course, property 5).

The single simplex in $\Sigma$ has $d$ vertices which we associate to
$x_{n-d+1},\ldots,x_n$ to simplify notation.  Geometrically this phase
corresponds to a orbifold $\C^d/G$, where $G$ is the finite
abelian group given by $N$ divided by the lattice generated by
the columns of $A$ associated to $x_{n-d+1},\ldots,x_n$.

The combinatorics of this phase is straight-forward. The Cox ideal and
Stanley--Reisner ideal are respectively:
\begin{equation}
\begin{split}
  B_\Sigma &= (x_1x_2\ldots x_r)\\
  I_\Sigma &= (x_1,x_2,\ldots, x_r).
\end{split}
\end{equation}
It follows that we have nonzero $D$-graded Betti numbers
$b_{0,\bbeta}$ for $\bbeta$ given by any of the first $r$ columns of
$\Phi$. Let us denote these vectors by
$\bbeta_1,\ldots,\bbeta_r\in D$.

The exact sequences (\ref{eq:LZN}) and (\ref{eq:M-D}) are split and so
we have an isomorphism of lattices
\begin{equation}
  L\oplus N \cong D\oplus M.    \label{eq:LNDM}
\end{equation}
Let us view this isomorphism as given by a $n\times n$ unimodular
integral matrix $C$. The first $r$ columns of $C$ are given by
${}^t\Phi$ and so the upper left $r\times r$ block of $C$ is given by
the vectors $\bbeta_1,\ldots,\bbeta_r$. Similarly the last $d$ columns
of $C^{-1}$ are given by ${}^tA$ and so the lower-right $d\times d$
block of $C^{-1}$ is given by the coordinates of
$x_{n-d+1},\ldots,x_n$. The determinant of this matrix is equal to
$|G|$. By the Schur complement this is also equal to the determinant
of the $r\times r$ matrix of vectors $\bbeta_1,\ldots,\bbeta_r$.

So, any $\balpha\in D$ may be written uniquely as
\begin{equation}
  \balpha = \sum_{i=0}^r t_k\bbeta_k,
\end{equation}
for rational numbers $t_k$. 
\begin{prop}
  The ``fundamental parallelepiped'' $0\leq t_k<1$ in $D$ contains
  $|G|$ vectors $\balpha_1,\balpha_2,\ldots$ which may be used 
  to generate $\DC(X_\Sigma)$.
\end{prop}
The fact that there are $|G|$ vectors follows from the statement above
about the determinant. 

Clearly $S/(x_j)$ is annihilated by $B_\Sigma$ for
$j=1,\ldots,r$. The short exact sequence
\begin{equation}
\xymatrix@1@M=2mm{
  0\ar[r]&S(-\bbeta_j+\bdelta)\ar[r]^-{x_j}&S(\bdelta)\ar[r]&
  {\displaystyle \frac{S}{x_j}(\bdelta)}\ar[r]&0,
}  \label{eq:sh1}
\end{equation}
gives an equivalence $\wS(-\bbeta_j+\bdelta)\cong \wS(\bdelta)$ in
$\DC(X_\Sigma)$. Hence we generate the whole of $\DC(X_\Sigma)$ from
the fundamental parallelepiped. \QED

Next we show that the local cohomology groups
$H^*_{B_\Sigma}(S)_\bdelta$ vanish for any $-\bdelta$ in the
fundamental parallelepiped. The resolution (\ref{eq:SRres}) is the
Koszul resolution of $(x_1,x_2,\ldots, x_r)$. Hence nonzero Betti
numbers $b_{i,\alpha}$ appear with vectors $\alpha$ with any
combination of 0's and 1's in the first $r$ positions and 0's in the
final $d$ positions. For any such $\alpha$ we may find a $v$ in the
kernel of $A$, as above, such that the positive entries of $v$
coincide with the 1's in $\alpha$ and again define $N_v$ as the sum of the
positive entries. Let $v$ be the image of $\mathbf{v}_{m,\Sigma}$ in
$L$. So, as before, the excluded region associated to $\alpha$ is
\begin{equation}
\begin{split}
  (-\bdelta,\mathbf{v}_{m,\Sigma}) &= \sum_i
  -t_i(\bbeta_i,\mathbf{v}_{m,\Sigma})\\
     &\leq -N_v.
\end{split}
\end{equation}
But $(\bbeta_i,\mathbf{v}_{m,\Sigma})$ is simply the $i$th entry in
the vector $v$. The inequality is therefore violated for $0\leq
t_i<1$.

It would be nice to show that the set of line bundles
$\wS(\balpha)$ for all $\balpha$ in the fundamental parallelepiped
form a tilting collection. This requires checking that the local
cohomology groups vanish for $\bdelta=\balpha_i-\balpha_j$. The
combinatorics of this is a little messy so we will content ourselves
with examples.

Suppose $d=3$ and the $n=7$ points of $\cA$ lie in the plane as:
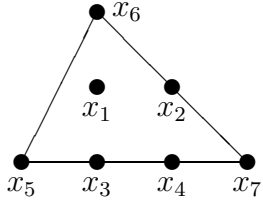
\unitlength=10mm
\begin{center}
\begin{picture}(3,2.1)
\put(0,0){\circle*{0.2}}
\put(0,0){\makebox(0,-0.2)[t]{$x_5$}}
\put(1,0){\circle*{0.2}}
\put(1,0){\makebox(0,-0.2)[t]{$x_3$}}
\put(2,0){\circle*{0.2}}
\put(2,0){\makebox(0,-0.2)[t]{$x_4$}}
\put(3,0){\circle*{0.2}}
\put(3,0){\makebox(0,-0.2)[t]{$x_7$}}
\put(1,1){\circle*{0.2}}
\put(1,1){\makebox(0,-0.2)[t]{$x_1$}}
\put(1,2){\circle*{0.2}}
\put(1.2,2){\makebox(0,0)[l]{$x_6$}}
\put(2,1){\circle*{0.2}}
\put(2,1){\makebox(0,-0.2)[t]{$x_2$}}
\put(0,0){\line(1,0){3}}
\put(0,0){\line(1,2){1}}
\put(1,2){\line(1,-1){2}}
\end{picture}
\end{center}
This may be written as
\begin{equation}
  A = \begin{pmatrix}1&1&1&1&1&1&1\\
  0&-1&0&-1&1&0&-2\\0&-1&-1&-2&0&1&-3\end{pmatrix},\quad
  \Phi= \begin{pmatrix}0&0&-2&-1&1&0&0\\0&0&1&-2&0&0&1\\
  1&-1&-1&1&0&0&0\\-2&0&1&0&0&1&0\end{pmatrix}
\end{equation}
and $X$ corresponds to an orbifold $\C^3/\Z_6$, where the $\Z_6$
action is generated by $(x_5,x_6,x_7)
\mapsto(e^{\frac{2\pi i}3}x_5,-x_6,e^{\frac{2\pi i}6}x_7)$.

The fundamental parallelepiped then contains the 6 points
\begin{equation}
\begin{split}
\balpha_1 &= (0,0,0,0)\\
\balpha_2 &= (0,1,0,0)\\
\balpha_3 &= (1,0,0,0)\\
\balpha_4 &= (0,0,0,1)\\
\balpha_5 &= (0,1,-1,1)\\
\balpha_6 &= (1,0,0,1).
\end{split}  \label{eq:Z6set}
\end{equation}
In this case one can explicitly check that the sum of the
corresponding six line bundles is a tilting sheaf.

The quiver associated to this tilting sheaf is, of course, nothing
other than the McKay quiver for this orbifold:
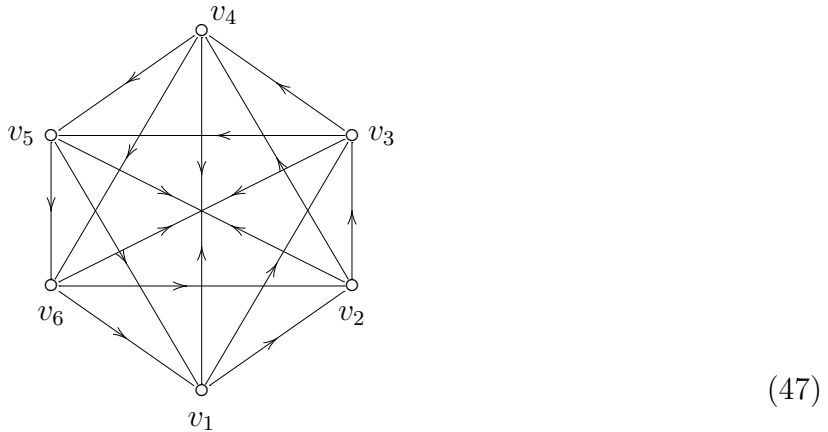
\begin{equation}
\begin{xy} <1.0mm,0mm>:
  (20,0)*{\circ}="a",(40,14)*{\circ}="b",(40,34)*{\circ}="c",
  (20,48)*{\circ}="d",(0,34)*{\circ}="e",(0,14)*{\circ}="f",
  (20,-4)*{v_1},(40,10)*{v_2},(44,34)*{v_3},
  (23,50)*{v_4},(-4,34)*{v_5},(0,10)*{v_6},
  \ar@{-}|*\dir{>}"a";"b"
  \ar@{-}|*\dir{>}"b";"c"
  \ar@{-}|*\dir{>}"c";"d"
  \ar@{-}|*\dir{>}"d";"e"
  \ar@{-}|*\dir{>}"e";"f"
  \ar@{-}|*\dir{>}"f";"a"
  \ar@{-}|*\dir{>}"a";"c"
  \ar@{-}|*\dir{>}"b";"d"
  \ar@{-}|(0.45)*\dir{>}"c";"e"
  \ar@{-}|*\dir{>}"d";"f"
  \ar@{-}|*\dir{>}"e";"a"
  \ar@{-}|(0.45)*\dir{>}"f";"b"
  \ar@{-}|(0.4)*\dir{>}|(0.6)*\dir{<}"a";"d"
  \ar@{-}|(0.4)*\dir{>}|(0.6)*\dir{<}"b";"e"
  \ar@{-}|(0.4)*\dir{>}|(0.6)*\dir{<}"c";"f"
\end{xy}  \label{eq:McKZ6}
\end{equation}

Returning to the general orbifold for a moment,
coherent sheaves on the stack $X_\Sigma=\C^d/G$ correspond to
$G$-equivariant coherent sheaves on $\C^d$. These in turn have
resolutions by $G$-equivariant bundles on $\C^d$ which are classified
by finite-dimensional representations of $G$ over $\C$. Obviously the
latter are generated by the $|G|$ one-dimensional irreducible
representations of the abelian group $G$. Indeed, the sheaves
$\wS(\balpha_i)$ we have found above correspond to these one-dimensional
representations. It follows that there can be no equivalences between
these generators in the derived category as then the rank of the
Grothendieck group would be wrong. So this unresolved phase is
wholesome.

What about the other phases? For this $\C^3/\Z_6$ orbifold there are a
total of 32 phases. That is, there are 32 triangulations of the point
set $\cA$ which all happen to be regular. The secondary polytope (see,
for example, \cite{BFS:secpol}) has 32 vertices each of which
corresponds to a phase. 5 of the phases are smooth resolutions and
there are 26 partial resolutions corresponding to the remaining
phases.

With a combination of Macaulay 2 (for computing the Betti numbers) and
Maple (for checking the required inequalities with the linear
programming package) it is not hard to show that in all 32 phases the
differences $\balpha_i-\balpha_j$ in (\ref{eq:Z6set}) are
$B_\Sigma$-acyclic.

We now need to check if the $\wS(\balpha_i)$'s generate the whole
derived category in every phase. This turns out to be a
combinatorially tricky question.  Let us return to the general
situation. We have a prime decomposition
\begin{equation}
  B_\Sigma = \bigcap_{k=1}^t \mf{m}_k,  \label{eq:pdecomp}
\end{equation}
where each $\mf{m}_k$ is a linearly generated
monomial ideal in $S$.
\begin{prop}
The subcategory $T_\Sigma$ of $\DC(\mathrm{gr-}S)$ is generated as a
triangulated subcategory by $S/\mf{m}_k$ (and its shifts) for all $k$.
\end{prop}
This proposition is very similar to proposition
\ref{prop:gen1}. Recall that $T_\Sigma$ is generated by modules
annihilated by $B_\Sigma^N$ for some $N$. Following the proof of
proposition \ref{prop:gen1} we may immediately see that $T_\Sigma$ is
generated by modules annihilated by $B_\Sigma$. So, suppose $M$ is
annihilated by $B_\Sigma$ and define
\begin{equation}
  M_{m} = \mf{m}_1\mf{m}_2\ldots\mf{m}_{m}M,
\end{equation}
and $M=M_0$. Then $M_t=0$ and we have a
short exact sequence
\begin{equation}
\xymatrix@1@M=2mm{
0\ar[r]&\mf{m}_mM_{m-1}\ar[r]&M_{m-1}\ar[r]&M'\ar[r]&0.
}
\end{equation}
Assume, by decreasing induction on $m$, that $M_m=\mf{m}_mM_{m-1}$ is
in the subcategory generated by $S/\mf{m}_k$ (and its shifts) for all
$k$.  Now $M'$ is annihilated by $\mf{m}_m$ and so is an
$(S/\mf{m}_m)$-module. Since $(S/\mf{m}_m)$ is a regular ring, we have
a finite free resolution of $M'$ in terms of $(S/\mf{m}_m)$ and its
grade-shifts. Therefore $M_{m-1}$ is
in the subcategory generated by $S/\mf{m}_k$ (and its shifts) for all
$k$.  \QED

\medskip

Let us write $\mf{m}=(x_{i_1},x_{i_2},\ldots,x_{i_p})$. We have a
Koszul resolution
\begin{equation}
\xymatrix@1@M=2mm{
  S(-\bbeta)\ar[r]&\ldots\ar[r]&\oplus_j
  S(-\boldsymbol{\Phi}_{i_j})\ar[r]&S\ar[r]&{\displaystyle\frac{S}{\mf{m}}}}.
\label{eq:relm}
\end{equation}
where $\boldsymbol{\Phi}_{i_j}$ is the $i_j$-th column of $\Phi$, and
$\bbeta$ is the sum of the columns over the index set
$\{i_1,i_2,\ldots, i_p\}$. Therefore, {\em all\/} of the relations between
line bundles in the derived category $\DC(X_\Sigma)$ are generated by
triviality of complexes of the form 
\begin{equation}
\xymatrix@1@M=2mm{
  \wS(-\bbeta)\ar[r]&\ldots\ar[r]&\oplus_j
  \wS(-\boldsymbol{\Phi}_{i_j})\ar[r]&\wS
},  \label{eq:relS}
\end{equation}
obtained from (\ref{eq:relm}) for $\mf{m}=\mf{m}_1,\ldots,\mf{m}_t$.

Now, to $\mf{m}=(x_{i_1},x_{i_2},\ldots,x_{i_p})$ appearing as a prime
factor of $B_\Sigma$ we may associate the corresponding generator 
$x_{i_1}x_{i_2}\ldots x_{i_p}$ of the Stanley--Reisner ideal
$I_\Sigma$. Therefore we have a nonzero Betti number $b_{0,\alpha}$ in 
(\ref{eq:betti}) where the 1's in $\alpha$ correspond to the locations 
$\{i_1,i_2,\ldots, i_p\}$. It immediately follows that the vector
$-\bbeta$ in (\ref{eq:relm}) is {\em not\/} $B_\Sigma$-acyclic.

The general idea of generating the whole derived category is to take a
vector $\bbeta$ which is not in our tilting set $\{\balpha_i\}$ and
use the relations (\ref{eq:relS}) (and their grade shifts) to replace
$\wS(\bbeta)$ by an isomorphic object in the derived category
represented by a complex of tilting objects. Combinatorially this is
messy and we will not try to confront this process directly.

Note also that the fact that $-\bbeta$ in (\ref{eq:relm}) is
guaranteed to not be $B_\Sigma$-acyclic seems to indicate that there
can be no relations in $\DC(X_\Sigma)$ within our tilting set.

Anyway, rather than attempting to prove that $\{\wS(\balpha_i)\}$ are
independent and generate $\DC(X_\Sigma)$ directly, we will resort to a
string theory argument. We know that $\{\wS(\balpha_i)\}$ are
independent objects in the orbifold phase. Therefore, these $|G|$
D-branes have central charges $Z(\wS(\balpha_i))$ which vary
holomorphically and independently over the orbifold phase. So,
therefore, they vary independently over the whole K\"ahler moduli
space. It follows that these D-branes are independent objects in
K-theory, and thus the Grothendieck group, and thus the derived
category in any phase. Furthermore, the K-theory group remains fixed
between phases (since it is the group of topological B-brane charges)
and so our set $\{\wS(\balpha_i)\}$ generates the whole derived
category.

So we arrive at the conclusion that $\{\wS(\balpha_i)\}$ are
independent and generate $\DC(X_\Sigma)$ in all 32 phases and so {\em
  this $\C^3/\Z_6$ example is wholesome}.

\subsection{An unwholesome example in dimension 5}  \label{ss:un5}

Consider $S=\C[x_0,\ldots,x_6]$ with charge matrix
\begin{equation}
  \Phi = \begin{pmatrix}-6&\pz1&\pz2&\pz1&\pz1&\pz1&\pz0\\
           \pz0&\pz1&-1&\pz0&\pz0&-1&\pz1\end{pmatrix}.
\end{equation}
This is yields a noncompact \CY\ fivefold with six phases. 

There is no tilting collection of free $S$-modules that works
simultaneously in all six phases but one may show that
\begin{equation}
\begin{split}
  T = \;&S(0,0)\oplus S(1,0)\oplus S(2,0)\oplus S(3,0)\oplus S(4,0)
        \oplus S(5,0)\oplus\\
    &S(0,1)\oplus S(1,1)\oplus S(2,1)\oplus S(3,1)\oplus S(4,1)
        \oplus S(5,1),
\end{split}
\end{equation}
satifies $\Ext_X^n(T,T)=0$ for $n\neq0$ everywhere except in one
phase.  Similarly
\begin{equation}
\begin{split}
  T = \;&S(0,0)\oplus S(1,0)\oplus S(2,0)\oplus S(3,0)\oplus S(4,0)
        \oplus S(5,0)\oplus\\
    S(-1,1)\oplus&S(0,1)\oplus S(1,1)\oplus S(2,1)\oplus S(3,1)\oplus S(4,1),
\end{split}
\end{equation}
satifies $\Ext_X^n(T,T)=0$ for $n\neq0$ everywhere except in one other
phase.

Actually this model has a $\Z_2$ symmetry $(x_0,\ldots,x_5)\to
(x_0,x_2,x_1,x_3,x_4,x_6,x_5)$. Dividing the moduli space out by this
symmetry we have only three inequivalent phases and then we {\em do\/}
have candidate tilting collections from above for all three phases. In this
sense this example can still actually be wholesome. However, the existence
of this example shows that the combinatorics of toric geometry do not
enforce wholesomeness.


\section{Relationship to $\Pi$-Stability}   \label{sec:pi-stab}

\subsection{The $\Z_3$-orbifold}  \label{ssec:z_3-orbifold}

So far we have been concerned with a simple description of the derived
category in terms of line bundles $\{\wS(\balpha_i)\}$ and the
resulting quiver. It appears, at least in many example, that the
description is constant over the whole K\"ahler moduli space in that
the same tilting set and quiver can be used in every phase. This
reflects the constancy of the B-model over the K\"ahler moduli space.

On the other hand, the way that our tilting collection generates the
derived category changes as we move from one phase to another since
all the combinatorics depend on the Stanley--Reisner ideal
$I_\Sigma$. Since whether a given object is $\Pi$-stable changes as
one moves around the moduli space, one might suspect that this is also
related to the same combinatorics. We will see in this section that
this is indeed the case.

It is perhaps easiest to begin with an example and then try to make
generalizations. Consider $\C^3/\Z_3$ for which $S=\C[p,x,y,z]$,
$d=3$, $r=1$ and the matrix $\Phi$ is given by $(-3\; 1\; 1\;
1)$. Following section \ref{ss:r=1}, the toric ideal is $(xyz-p^3)$.
Note that we will restrict attention to the ``physicists'' notion of
$\Pi$-stability where one studies how stability varies over the moduli
space of complexified K\"ahler forms. The more general notion of the
full space of stability conditions was studied by Bridgeland in this
same example in \cite{bridge:t-loc,Bridge:Z3}.

The phase $\Sigma_-$ corresponds to a Stanley--Reisner ideal $(p)$ and
yields the orbifold phase. The phase $\Sigma_+$ corresponds to a
Stanley--Reisner ideal $(xyz)$ and corresponds to the geometry of
$\O_{\P^2}(-3)$ which is the ``large radius'' resolved phase.

The tilting set in both phases is
\begin{equation}
  \{\wS,\wS(1),\wS(2)\}
\end{equation}
and the quiver is given in (\ref{eq:egO(-3)}).

Inspired by the observation in \cite{HHP:linphase} that the three
D-branes in this set are somehow ``globally defined'' over the whole
K\"ahler moduli space in terms of the gauged linear sigma model we
would like to propose that {\em this set of three D-branes is
  everywhere $\Pi$-stable}.

Recall that $\Pi$-stability is governed by the
phase\footnote{Sometimes the word ``grade'' is used instead but we
  already have another notion of grade here.} of a D-brane $\cF$:
\begin{equation}
  \xi(\cF) = \frac1{\pi}\arg Z(\cF),  \label{eq:phase}
\end{equation}
where $Z$ is the central charge. Of course, one must be careful about
defining the mod 2 ambiguity in (\ref{eq:phase}). In our case the
D-branes $\wS(a)$ have noncompact support and thus infinite central
charge. In this case, the phase is defined purely by the dimension of
the support (see, for example, section 6.2.5 of \cite{me:TASI-D}):
\begin{equation}
\begin{split}
  \xi(\wS(a)) &= -\ff12\dim(X_\Sigma), \quad\hbox{for all $a\in\Z$.}\\
    &= -\ff32.
\end{split}
\end{equation}

Now consider $\cE=\O_{\P^2}$, the structure sheaf of the exceptional
$\P^2$. The locus of the exceptional divisor in $X_{\Sigma_+}$ is
given by $p=0$ and so we have an exact sequence:
\begin{equation}
\xymatrix@1@M=2mm{
  0\ar[r]&\wS(3)\ar[r]^-p&\wS\ar[r]&\cE\ar[r]&0,
}  \label{eq:seqE}
\end{equation}
or, to write in terms of a triangle:
\begin{equation}
\xymatrix{
&\cE\ar[dl]|{[1]}&\\
\wS(3)\ar[rr]^p&&\wS\ar[ul].
}  \label{eq:tri1}
\end{equation}

It is known that $\cE$ is massless at the ``conifold point''
where the CFT becomes singular \cite{DFR:orbifold}. This implies that
near this conifold point, the phase of $\cE$ can take on any
value. This, in turn, implies that the grade difference on the left or
right edge of (\ref{eq:tri1}) can exceed one causing a decay of the
opposite vertex. What actually happens is sketched in figure
\ref{fig:Z3marg}
obtained by numerical integration of the Picard--Fuchs equation as in
section 7.3 of \cite{me:TASI-D}.

\iffigs
\begin{figure}
\begin{center}
\setlength{\unitlength}{0.00050000in}%
\begin{picture}(10566,5766)(868,-6094)
\thinlines
\put(7201,-4861){\circle*{150}}
\put(9601,-4861){\circle*{150}}
\put(4801,-4861){\circle*{150}}
\put(2401,-4861){\circle*{150}}
\thicklines
\put(6001,-6061){\vector( 0, 1){5700}}
\put(901,-5761){\vector( 1, 0){10500}}
\thinlines
\put(4801,-4861){\line( 0, 1){4500}}
\put(7201,-4861){\line( 0, 1){4500}}
\put(9601,-361){\line( 0,-1){4500}}
\put(2401,-4861){\line( 0, 1){4500}}
\put(4801,-4861){
}
\put(2401,-4861){
}
\put(7201,-4861){
}
\put(9601,-4861){
}
\put(7201,-4861){
}
\put(4801,-4861){
}
\put(7201,-4861){
}
\put(9613,-4861){
}
\put(4802,-4869){
}
\put(7203,-4869){
}
\put(2401,-4869){
}
\put(4783,-4871){
}
\put(11251,-6061){\makebox(0,0)[lb]{\smash{$B$}}}
\put(5701,-586){\makebox(0,0)[lb]{\smash{$J$}}}
\put(9726,-5086){\makebox(0,0)[lb]{\smash{$\cE(1)$}}}
\put(1751,-5361){\makebox(0,0)[lb]{\smash{$\cE(-2)$}}}
\put(4651,-5311){\makebox(0,0)[lb]{\smash{$\cE(-1)$}}}
\put(7101,-5336){\makebox(0,0)[lb]{\smash{$\cE$}}}
\put(7201,-5761){\circle*{100}}
\put(7150,-6200){\makebox(0,0)[lb]{\smash{$\ff12$}}}
\put(4801,-5761){\circle*{100}}
\put(4600,-6200){\makebox(0,0)[lb]{\smash{$-\ff12$}}}
\put(6200,-4100){\makebox(0,0)[lb]{\smash{$P_0$}}}
\put(6010,-1830){\circle*{100}}
\put(6200,-1910){\makebox(0,0)[lb]{\smash{$P_1$}}}
\put(6010,-4020){\circle*{100}}

\end{picture}
\end{center}
  \caption{Lines of marginal stability for $\C^3/\Z_3$.}
  \label{fig:Z3marg}
\end{figure}
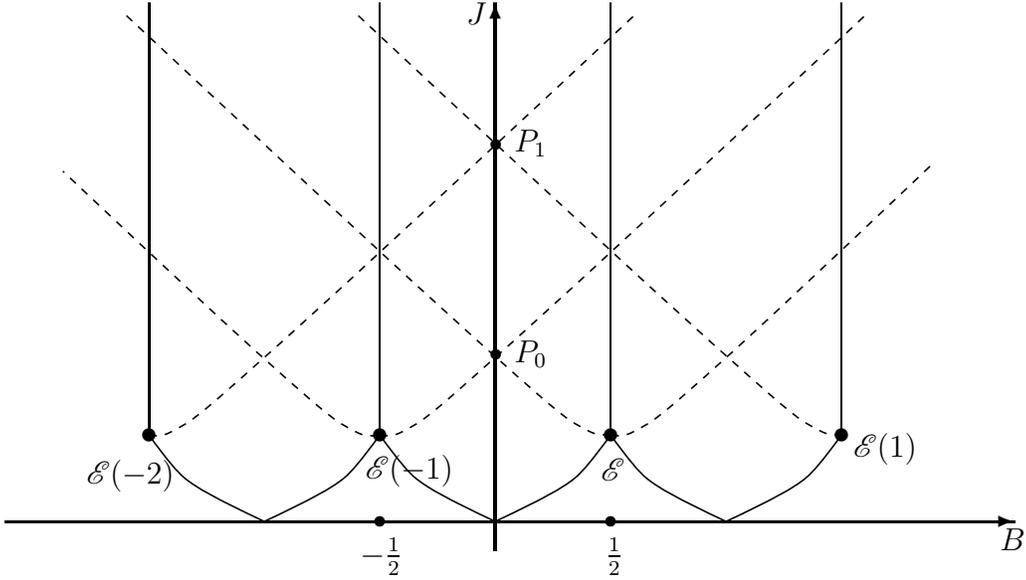
\fi

The figure is read as follows. The sketch is of the $(B+iJ)$-plane
where $J$ is the K\"ahler form. The solid lines denote the boundaries
of ``fundamental regions'' of the moduli space viewing the
$(B+iJ)$-plane as a Teichm\"uller space, roughly speaking. We have a
copy of the moduli space between $B=-\ff12$ and $\ff12$ where the
region is squeezed to width less than one as one approaches the
orbifold point at $B=J=0$. At the conifold point we have our massless
D-brane $\cE$. Actually whether one considers this to be $\cE$ or
$\cE(-1)$ depends upon paths taken in the moduli space. We have two
points in the $B+iJ$ plane denoted by dots in the figure corresponding the
where these D-branes become massless. Similarly $\cE(-2)$ and $\cE(1)$
become massless if one follows paths to other copies of the moduli
space.

The dashed lines in the figure represent lines of marginal stability
which are relevant to this discussion. Asymptotically, for large
$|B|$, these lines become straight and at an angle of $45^\circ$ to
the $B$-axis.

Consider starting at large radius limit, high on the $J$-axis.  Since
$\wS(a)$ is a $\mu$-stable line bundle for any $a\in\Z$, we expect
this to correspond to a $\Pi$-stable D-brane for sufficiently large
$J$.  As one moves down, one eventually reaches the point labeled $P_0$ in
the figure. At this point one hits the line of marginal stability
radiating leftwards out of the massless $\cE$ conifold point. At this
instant, the grade of $\cE$ rises above $-\ff12$. This causes $\wS(3)$
to decay in (\ref{eq:tri1}). Thus $\wS(3)$ is unstable as one nears
the orbifold phase.

Similarly consider the triangle:
\begin{equation}
\xymatrix{
&\cE(1)\ar[dl]|{[1]}&\\
\wS(4)\ar[rr]^p&&\wS(1)\ar[ul].
} 
\end{equation}
As we see from figure \ref{fig:Z3marg}, the marginal line coming left
out of the $\cE(1)$ conifold point also crosses the $J$-axis which
causes the decay of $\wS(4)$ at $P_1$. Similarly all D-branes of the form
$\wS(a)$ for $a\geq 3$ decay as one moves into the orbifold phase. The
larger the value of $a$, the larger the value of $J$ at which the
decay takes place.

Now consider the triangle
\begin{equation}
\xymatrix{
&\cE(-1)\ar[dl]|{[1]}&\\
\wS(2)\ar[rr]^p&&\wS(-1)\ar[ul].
} 
\end{equation}
As one moves down the $J$-axis and hits $P_0$, the grade of
$\cE(-1)$ {\em falls below\/} $-\ff32$. This causes a decay of
$\wS(-1)$. Similarly all D-branes of the form
$\wS(a)$ for $a<0$ decay as one moves into the orbifold phase.

This yields the result that {\em the only D-branes of the form
  $\wS(a)$ which remain stable as one moves from the resolved phase to
  the orbifold phase are those in the tilting set.}
Furthermore, we have a rather explicit picture of how this
happens. For large positive $a$, $\wS(a)$ iteratively decays into $\cE(a')$ plus
$\wS(a')$ with $a'=a-3$ until $a$ falls below 3. Similarly for
negative $a$, the decay increases $a$ by 3 until it is positive.

It is instructive to describe the D-brane, $\cE$, which is massless at
the conifold point in terms of a quiver representation. The sequence
(\ref{eq:seqE}) expresses $\cE$ in terms of $\wS$ and $\wS(3)$. But we
know in the resolved phase that $\wS$ can be expressed in terms of the
tilting set $\{\wS,\wS(1),\wS(2)\}$. We may rewrite (\ref{eq:seqE}) as
\begin{equation}
\xymatrix@1@M=2mm{
  0\ar[r]&\wS\ar[r]&\wS(1)^{\oplus3}\ar[r]&
    \wS(2)^{\oplus3}\ar[rr]^-{(px,py,pz)}&&\wS\ar[r]&\cE\ar[r]&0.
}  \label{eq:seqE2}
\end{equation}
In other words, $\cE$ is the cokernel of the map $(px,py,pz)$. In
terms of quivers, we know that the free module $S$ corresponds to the
infinite set of paths ending at the node $v_0$ in
(\ref{eq:egO(-3)}). Any such path that is not of length zero must end
in $px$, $py$ or $pz$. Thus, the module corresponding to the cokernel
of the map $(px,py,pz)$ is precisely the one-dimensional quiver
representation with the single dimension associated to vertex $v_0$. 
This ``simple'' representation of the path algebra is familiar as the
fractional brane which becomes massless at the conifold point
\cite{DFR:orbifold}.

One may try to picture a similar effect starting in the orbifold phase
and moving into the resolved phase. However, the result is not as
pretty since we cannot begin with the assumption that the $\wS(a)$'s
are all stable at the orbifold point.

There is one ``symmetry'' between the resolved phase and the
orbifold phase which is worth emphasizing. In the resolved phase
corresponding to the triangulation $\Sigma_+$, we have
$B_{\Sigma_+}=(x,y,z)$. In the orbifold phase $X_{\Sigma_-}$ we have
$B_{\Sigma_-}=(p)$. In the resolved phase we may write the D-brane, $\cE$,
which becomes massless at the conifold point as $S/B_{\Sigma_-}$. 
In the orbifold phase we have $\wS(3)\cong\wS$, which, when applied to 
(\ref{eq:seqE2}) yields $\cE\cong S/B_{\Sigma_+}(3)$.

So, in the resolved phase, $S/B_{\Sigma_+}$ (and its grade-shifts)
play the role of a ``no-brane'' while $S/B_{\Sigma_-}$ is the brane
massless at the conifold. In the orbifold phase the r\^oles are
reversed.

\subsection{The conifold}  \label{ssec:conifold-S}

The conifold of section \ref{ss:r=1} is a little less
satisfying. Assume we are in the phase given by $B_\Sigma=(x,y)$. The
massless D-brane of interest is therefore given by the module
$S/(z,w)$. This corresponds to the structure sheaf $\O_C$ of the
exceptional curve $C$ in the small resolution. We have a resolution:
\begin{equation}
\xymatrix@1@M=2mm{
0\ar[r]&\wS(2)\ar[r]^-{\left(\begin{smallmatrix}-w\\z
   \end{smallmatrix}\right)}
&\wS(1)\oplus\wS(1)\ar[r]^-{(z\;w)}&\wS\ar[r]
&\O_C\ar[r]&0.
}
\end{equation}
Write this as a triangle
\begin{equation}
\xymatrix{
&\O_C[-1]\ar[dl]|{[1]}&\\
\wS(2)\ar[rr]&&\cX\ar[ul].
} \label{eq:tri-con}
\end{equation}
where
\begin{equation}
\cX = \xymatrix@1@M=2mm{
  \poso{\wS(1)\oplus\wS(1)}\ar[r]&\wS.
} 
\end{equation}

Clearly the phase of $\wS(a)$ is always $-\ff32$ for any $a\in\Z$ as
in the previous section. Similarly $\cX$ has phase $-\ff32$ everywhere
in the moduli space. The phase of $\O_C[-1]$ at large radius is also
$-\ff32$ which renders $\wS(2)$ stable, as one would expect. Now as we
follow the $J$-axis down the grade of $\O_C[-1]$ starts to rise and so
we might hope that it eventually increases above $-\ff12$ to
destabilize $\wS(2)$. What actually happens is shown in figure \ref{fig:con}.

\iffigs
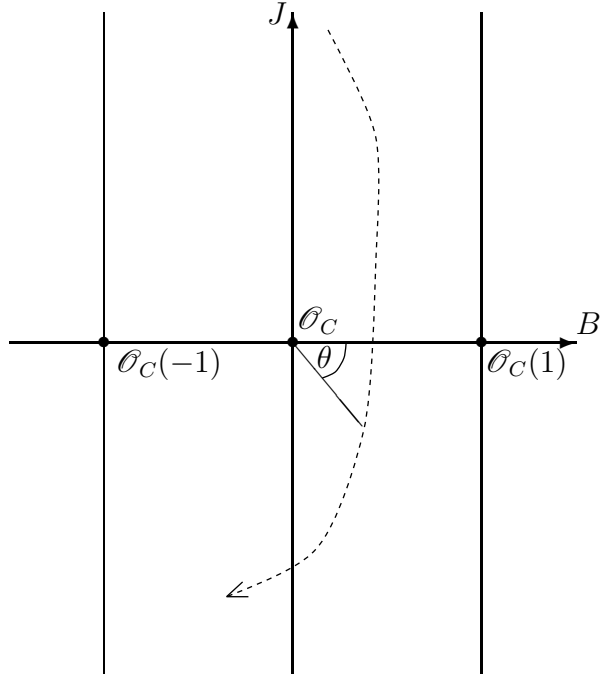
\begin{figure}
\begin{center}
\setlength{\unitlength}{0.00041700in}%
\begin{picture}(7266,8466)(1168,-8794)
\thinlines
\put(4801,-4561){\circle*{150}}
\put(7201,-4561){\circle*{150}}
\put(2401,-4561){\circle*{150}}
\thicklines
\put(4801,-8761){\vector( 0, 1){8400}}
\put(1201,-4561){\vector( 1, 0){7200}}
\thinlines
\put(7201,-361){\line( 0,-1){8400}}
\put(2401,-361){\line( 0,-1){8400}}
\put(4801,-4561){\line( 5,-6){885.246}}
\put(5251,-586){
}
\put(5176,-5011){
}
\put(4476,-511){\makebox(0,0)[lb]{\smash{$J$}}}
\put(8400,-4460){\makebox(0,0)[lb]{\smash{$B$}}}
\put(4851,-4390){\makebox(0,0)[lb]{\smash{$\O_C$}}}
\put(7251,-4930){\makebox(0,0)[lb]{\smash{$\O_C(1)$}}}
\put(2551,-4936){\makebox(0,0)[lb]{\smash{$\O_C(-1)$}}}
\put(5108,-4881){\makebox(0,0)[lb]{\smash{$\theta$}}}
\end{picture}
\end{center}
  \caption{Decay for a conifold.}
  \label{fig:con}
\end{figure}
\fi

A fundamental region now looks like a vertical strip of infinite
length between $B=0$ (where $\O_C$ is massless) and $B=1$ (where
$\O_C(1)$ is massless). The phase ``boundary'' is the line $J=0$ which
separates the two \CY\ phases related by a flop.
It is not hard to show (see section 7.2 of
\cite{me:TASI-D} for example) that 
\begin{equation}
  \xi(\O_C[-1]) = -1 +\frac{\theta}{\pi},
\end{equation}
where $\theta$ is the angle shown in figure \ref{fig:con}. It follows
that so long as we stay in this fundamental strip, the D-brane $\wS(2)$
never decays via the triangle (\ref{eq:tri-con}). Only when one
reaches the large radius limit of the flopped \CY\ when $\theta=\pi/2$
does $\wS(2)$ become marginally unstable. Alternatively, if one
ventures into the neighboring phase to the left, as shown by the
dotted path in the figure, the bundle $\wS(2)$ does decay.

We have therefore demonstrated that $\wS(2)$ is unstable ``in a way''
when one ventures into the flopped phase, but only when one actually
reaches the large radius limit of the flop, or if one winds
sufficiently far around the conifold point in the moduli space to
enter another fundamental region.

Similar remarks also apply to all the other line bundles $\wS(a)$ for
$a$ anything other than 0 or 1. So again we have the result that the
tilting set $\{\wS,\wS(1)\}$ is somehow globally stable (not crossing
the walls of the fundamental region for the phase $J>0$) in the moduli
space while the other line bundles $\wS(a)$ are not.

The essential difference between the orbifold of section
\ref{ssec:z_3-orbifold} and this conifold is the codimension of the
exceptional set as we will see in the next section.

\subsection{A general picture}

Let us try to make some general comments about $\Pi$-stability based
on the above examples. The idea is that we will assume that all
``interesting'' decays are based on triangles coming from the kinds of
resolutions we have seen so far. In particular, in any phase $\Sigma$,
we have the Cox ideal $B_\Sigma$ with its primary
decomposition (\ref{eq:pdecomp}). The Alexander dual to this statement
is that the Stanley--Reisner ideal can be written
\begin{equation}
  I_\Sigma = \mf{m}_1^\vee+\mf{m}_2^\vee+\ldots+\mf{m}_t^\vee,
\end{equation}
where each $\mf{m}_i^\vee$ is a principal ideal. The only triangles we
concern ourselves with are the Koszul resolutions of $\mf{m}_i$ where
we consider all such primary ideals from all phases $\Sigma$. So all
our statements about $\Pi$-stability will be limited in the sense that
only a subset of all distinguished triangles are used.

Suppose we have a point set $\cA$ which is {\em wholesome} and we
choose a tilting set $\{S(\balpha_i)\}$. No line bundle is globally
stable over the whole K\"ahler moduli space for arbitrary paths but we
may make the situation more manageable by making cuts. That is, we fix
a fundamental region of the Teichm\"uller space much as in figure
\ref{fig:Z3marg}. We remove from consideration paths that cross the
walls of this fundamental region.  Now let us boldly assert that {\em
  there is a choice of cuts such that every object in the tilting set
  is stable over the whole moduli space}.

This is very similar in spirit to the picture in
\cite{HHP:linphase}. There they showed that only D-branes that lived
within a certain grade-restricted window could be ``globally defined''
over the whole moduli space of gauged linear sigma models. Actually
our assertion does not quite coincide with the analysis of
\cite{HHP:linphase}. Using the simplest ansatz for A-branes on the
Coulomb branch, the authors of \cite{HHP:linphase} were able to give
an example where the tilting set was {\em not\/} globally
defined. Hopefully this discrepancy can be avoided by using more
subtle A-branes.

In general there may be many large radius \CY\ phases.  In any such
large radius limit we expect line bundles $\wS(\bdelta)$ to be stable
for all possible $\bdelta\in D$. Choose such a \CY\ phase and denote
it by
$\Sigma_\infty$ and the associated Cox ideal by $B_\infty$. Consider some
other phase $\Sigma$ with a prime decomposition of $B_\Sigma$ given by
(\ref{eq:pdecomp}). To each prime ideal $\mf{m}$ in this decomposition
there is a Koszul resolution given by (\ref{eq:relm}) which we write
again for convenience:
\begin{equation}
\xymatrix@1@M=2mm{
  S(-\bbeta)\ar[r]&\ldots\ar[r]&\oplus_j
  S(-\boldsymbol{\Phi}_{i_j})\ar[r]&S\ar[r]&{\displaystyle\frac{S}{\mf{m}}}}.
\label{eq:relm2}
\end{equation}
To this we can associate two distinguished triangles:
\begin{equation}
\xymatrix@C=2cm{
&{\displaystyle\frac{S}{\mf{m}}}\ar[dl]|{[1]}&\\
\Bigl(S(-\bbeta)\to\ldots\to\oplus_j
  S(-\boldsymbol{\Phi}_{i_j})\Bigr)\ar[rr]&&S\ar[ul].
} \label{eq:gend1}
\end{equation}
and
\begin{equation}
\xymatrix@C=2cm{
&{\displaystyle\frac{S}{\mf{m}}[1-c]}\ar[dl]|{[1]}&\\
S(-\bbeta)\ar[rr]&&\Bigl(\to\ldots\to\oplus_j
  S(-\boldsymbol{\Phi}_{i_j})\to S\Bigr)\ar[ul],
} \label{eq:gend2}
\end{equation}
where $c$ is the codimension of the ideal $\mf{m}$ in $S$.

Suppose $\mf{m} \supset B_\infty$. In this case $S/\mf{m}$ is
annihilated by $B_\infty$ and so corresponds to a no-brane in the
large radius phase. The triangles above express relations between
$S(\bbeta)$'s.

Now suppose $\mf{m} \not\supset B_\infty$. In this case $S/\mf{m}$ is
{\em not\/} annihilated by $B_\infty$ and so corresponds to a
non-trivial brane in the large radius phase $X_\Sigma$. Now the above triangles
(and their grade shifts) express possible decay paths which can
destabilize various $\wS(\bbeta)$'s. Note first that our assumption
that the bundles in the tilting set $\{\wS(\balpha_i)\}$ are always
stable is entirely consistent with these triangles. The statement that
the tilting set is wholesome means that the elements of the tilting
set are independent and therefore we can never write a triangle of the
forms (\ref{eq:gend1}) or (\ref{eq:gend2}) (or their grade shifts)
expressing a decay of one tilting element into a combination of the
others and modules of the form $S/\mf{m}$.

Whether or not we actually have a decay of one line bundle into other
line bundles depends on the analysis of the phases. In triangles
(\ref{eq:gend1}) or (\ref{eq:gend2}) the objects at the bottom left
and bottom right of the triangle both always have phase $-d/2$ since
they correspond to bundles supported over the entire space. The only
phase which varies is therefore that of $S/\mf{m}$. If $S/\mf{m}$
becomes massless somewhere in the moduli space then we are in the
situation similar to sections \ref{ssec:z_3-orbifold} and
\ref{ssec:conifold-S}. 

At large radius limit, the sheaf associated to $S/\mf{m}$ has phase
$-\ff12(d-c)$. So, for example, the phase difference on the right edge
of the triangle (\ref{eq:gend2}) is $1-c/2$. In order to cause a decay
of $S(-\bbeta)$, this difference must rise to 1.  Let us assume we
follow a path that runs very close to the ``conifold point'' where
$S/\mf{m}$ becomes massless. We also assume that $Z(S/\mf{m})$ has a
simple zero at this point.  Then for decay, the path from the large radius
limit needs to subtend an angle of $\pi c/2$ with respect to this
point. For section \ref{ssec:z_3-orbifold} we had $c=1$ and so we only
needed to pass through an angle of $\pi/2$ which happened as we passed
from one phase to another. In section \ref{ssec:conifold-S}, for the
flop, we had an angle of $\pi$ which required going all the way to the
limit of the other phase. Clearly, for higher-dimensional examples
where $c>2$ we need to start to loop around the conifold point to get
the $\wS(a)$'s to decay.

So, if $\mf{m}$ has codimension one (i.e., is a
principal ideal) one would expect the analysis of section
\ref{ssec:z_3-orbifold} to follow and some line bundles outside the
tilting set will decay as we move from phase $\Sigma_\infty$ to phase
$\Sigma$. If the codimension of $\mf{m}$ is greater than one then this
need not happen and one would need to work harder, by looping around
in the moduli space, to see the non-tilting line bundles decay.


\section{Discussion}

We have seen how, in some examples, one may define a tilting set of
line bundles which works globally over the whole moduli space. Thus,
the derived category is given by the same quiver in each phase and we
recover the result that $\DC(X)$ is invariant in a very explicit way.
The example of section \ref{ss:un5} shows that we cannot always expect
such a global set to exist but this wholesomeness does seem
surprisingly ubiquitous in examples studied.  Given the usefulness of
such wholesome tilting sets it would be nice to find the precise
combinatorics of when they exist.

The combinatorial problems we encounter are classic in
combinatorial commutative algebra, such as analysis of
Stanley--Reisner ideals and local cohomology. One might therefore hope
that this well-developed branch of mathematics might offer some tools
and techniques that can extend the results of this paper.

We also have a rather paltry understanding of $\Pi$-stability at
present. What one would really like to know is, given a point in the
K\"ahler moduli space, what is the precise set of $\Pi$-stable objects in
$\DC(X)$. Viewing the derived category in terms of tilting line
bundles seems to offer some handle on this difficult problem, as we
saw in section \ref{sec:pi-stab}, but obviously much remains to be understood.


\section*{Acknowledgments}

I wish to thank W.~Allard, A.~Craw, D.~Eisenbud, R.~Karp, E.~Miller,
D.~Morrison, R.~Plesser and A.~Roy for useful discussions.  The author
is supported by an NSF grant DMS--0606578.


\end{document}